\documentclass[twocolumn,10pt]{elsarticle}

\usepackage{amssymb}
\usepackage{amsmath}

\usepackage{lineno}
\usepackage{graphicx}
\usepackage{dcolumn}
\usepackage{bm}
\usepackage{xcolor}
\usepackage[a4paper, top=2cm, bottom=2cm, left=2.5cm, right=2.5cm]{geometry}
\usepackage{comment}

\journal{Small}

\begin{document}

\onecolumn
\begin{frontmatter}



\title{Soft Colloidal Robots: Magnetically Guided Liquid Crystal Torons for Targeted Micro-Cargo Delivery}


\author[inst1,inst2]{Joel Torres}

\affiliation[inst1]{organization={Departament de Ciència de Materials i Química Física, Universitat de Barcelona},
            addressline={Martí i Franquès 1}, 
            city={Barcelona},
            postcode={08028}, 
            state={Catalonia},
            country={Spain}}
            
\affiliation[inst2]{organization={Institute of Nanoscience and Nanotechnology, IN2UB, Universitat de Barcelona},
            addressline={Martí i Franquès 1}, 
            city={Barcelona},
            postcode={08028}, 
            state={Catalonia},
            country={Spain}}

\author[inst3,inst4]{Rodrigo C. V. Coelho}

\affiliation[inst3]{organization={Centro Brasileiro de Pesquisas Físicas},
            addressline={Rua Xavier Sigaud 150}, 
            city={Rio de Janeiro},
            postcode={22290-180},
            country={Brazil}}
 
\affiliation[inst4]{organization={Centro de Física Teórica e Computacional, Faculdade de Ciências, Universidade de Lisboa},
            city={Lisboa},
            postcode={1749-016},
            country={Portugal}}
            
\author[inst5]{Patrick Oswald}

\affiliation[inst5]{organization={Université de Lyon, ENS de Lyon, Université Claude Bernard, CNRS, Laboratoire de Physique },
            city={Lyon},
            postcode={69342},
            country={France}}

\author[inst1,inst2]{Francesc Sagués}
\author[inst1,inst1]{Jordi Ignés-Mullol}

\begin{abstract}

Quasiparticles in liquid crystals, such as torons and skyrmions, represent a new class of topologically protected solitonic excitations, offering a promising route toward soft microrobotics. Here we demonstrate that torons can be propelled by modulated electric fields and magnetically steered with full directional control, thus achieving programmable trajectories without net liquid flow. Within microfluidic architectures, we guide ensembles of torons through confined channels and realize targeted pick-up, transport, and release of colloidal cargo. By combining experiments and numerical simulations, we uncover how magnetic alignment reshapes toron structure, speed, and stability, while confinement within microchannels gives rise to novel transport regimes. Unlike conventional colloidal inclusions, torons are intrinsically uniform, soft, and reconfigurable, establishing them as both an ideal model system for studying emergent phenomena in active topological matter and a versatile platform for next-generation soft robots, adaptive delivery systems, and smart active matter.
\end{abstract}




\end{frontmatter}


\twocolumn

\makeatletter
\newcommand{\manuallabel}[2]{\def\@currentlabel{#2}\label{#1}}
\makeatother

\manuallabel{SFig:EField}{S1}
\manuallabel{SFig:Yjunction_exp}{S2}
\manuallabel{SFig:anchoring}{S3}
\manuallabel{SFig:Expvelstraightwall}{S4}
\manuallabel{SFig:Yjunction_exp_comparison}{S5}

\manuallabel{SMov:skyrmion_H_exp}{S1}
\manuallabel{SMov:twoTICs}{S2}
\manuallabel{SMov:skyrmion_H_sim}{S3}
\manuallabel{SMov:transport}{S4}
\manuallabel{SMov:laser}{S5}
\manuallabel{SMov:funnel}{S6}
\manuallabel{SMov:sim_funnel}{S7}
\manuallabel{SMov:column}{S8}
\manuallabel{SMov:wall}{S9}

\section{Introduction}
\label{sec:introduction}


Active colloids are fascinating soft matter systems that can convert energy from the environment into non-conservative forces at the single particle level, typically a result of tailored external fields or physico-chemical interactions with particle surroundings \cite{Bishop2023,Yan2022,Bunea2020}.  Different mechanism have been demonstrated to drive either individual particles or swarms \cite{Sharan2021}.
Propulsion can originate from chemical reactions on the particle surface \cite{Howse2007,Hong2007}, or induced externally due to electric \cite{Bricard2013,Gangwal2008} or magnetic fields \cite{Tierno2007,Liao2019}, light sources \cite{Lozano2016, Arya2020}, or even temperature gradients \cite{Piazza2008,kolacz2020}. Active colloids have been the subject of intense multidisciplinary research: experimental, as well as theoretical or numerical, for the past two decades. Besides their fundamental interest, they have also been explored for the removal of water pollutants \cite{Katuri2017} or in drug delivery systems \cite{Bunea2020, XU2022121551}, with strong implications in the incipient domain of colloidal robots, which add addressability and on-time response to the self-propulsion \cite{WANG2013531,Ju2025}. Additionally, active colloids serve as useful models for mimicking complex physical and biological systems \cite{Doostmohammadi2018,Bunea2020,Fu2017}.

Regardless of the underlying physicochemical mechanism, propulsion relies on one basic principle: breaking the particle's fore-aft symmetry \cite{C0SM00953A, Yeomans2014}. When dispersed in Newtonian fluids, which encompass the vast majority of past and recent research, this asymmetry must be provided by the particles themselves. The situation can change for particles embedded in a non-Newtonian fluid \cite{Ignes-Mullol2022}. In that case, the fore-aft symmetry may be broken by the very nature of the fluid. A remarkable example is that of colloidal entities dispersed in liquid crystals (LCs) \cite{Smalyukh2025LiquidColloids}, which are intrinsically anisotropic fluids, that can be driven and oriented using external electromagnetic fields. For instance, the propulsion of spherical particles has been achieved within nematic LCs using oscillating electric fields due to the emergence of asymmetric ionic flows, which result in particles moving along the direction of the local nematic director field \cite{Peng2019,Lavrentovich2016}.   

Within this context, the concept of active or driven quasiparticles has emerged as a new topic in active colloidal matter. Here, the colloidal particles are substituted by topological solitonic structures in the LC orientational field \cite{Fumeron2023}. These structures generate spontaneously after the relaxation  of an electrohydrodynamic instability or during transition from the isotropic to the chiral nematic phase. In general, they appear due to a topological discordance between the boundary conditions and the bulk alignment \cite{OswaldBook, Doi2013, Kleman2003}. Unlike LC defects, which are generally unstable and tend to annihilate in complementary pairs \cite{Dierking2005AnnihilationFields}, some solitonic distortions can remain stable because the LC orientational field cannot relax without crossing costly elastic energy barriers. 

In recent years, a number of such topologically-protected structures have been identified \cite{Wu2022}. Some of them, particularly those that form in frustrated (strongly-confined) chiral nematic LC can be propelled within the LC layer by means of modulated alternating current (AC) electric fields \cite{Sohn2018,Ackerman2017, Shen2022RecentCrystals}. This intervention breaks the fore-aft symmetry of otherwise symmetric structures, generating net propulsion perpendicular to the surrounding LC director field
. This particle-like behavior is enriched by the fact that the same electric field can be used to tune their long and short range interactions, from attractive to repulsive \cite{Sohn2019}. Besides local propulsion, microfluidic applications demand addressability and reconfigurable control of the direction of motion of the colloidal quasiparticles. A possibility, used in the past to steer colloidal particles, is using a photosensitive anchoring layer on the confining surfaces to control the orientation of the surrounding LC molecules \cite{Hernandez-Navarro2014}. Since the stability of LC quasiparticles strongly depends on the boundary conditions of the LC layer  \cite{Tai2020}, alternative steering methods must be proposed, based on directly addressing the bulk LC orientation.

In this work, we propose a novel mechanism for steering a type of LC quasiparticles called torons based on controlling the surrounding LC director field through the application of a permanent magnetic field, thus leveraging the magnetic susceptibility anisotropy of the LC molecules. We highlight the benefits of LC quasiparticles over traditional colloidal inclusions and we demonstrate the capabilities of LC torons  for microfluidic applications, demonstrating their potential as micro-cargo transporters and analyzing their behavior when driven within microchannels.

\section{Materials and methods}

\subsection{Assembly of liquid crystal cells}
\label{sec:assembly}

\begin{figure}[t]
\center
\includegraphics[width=\columnwidth]{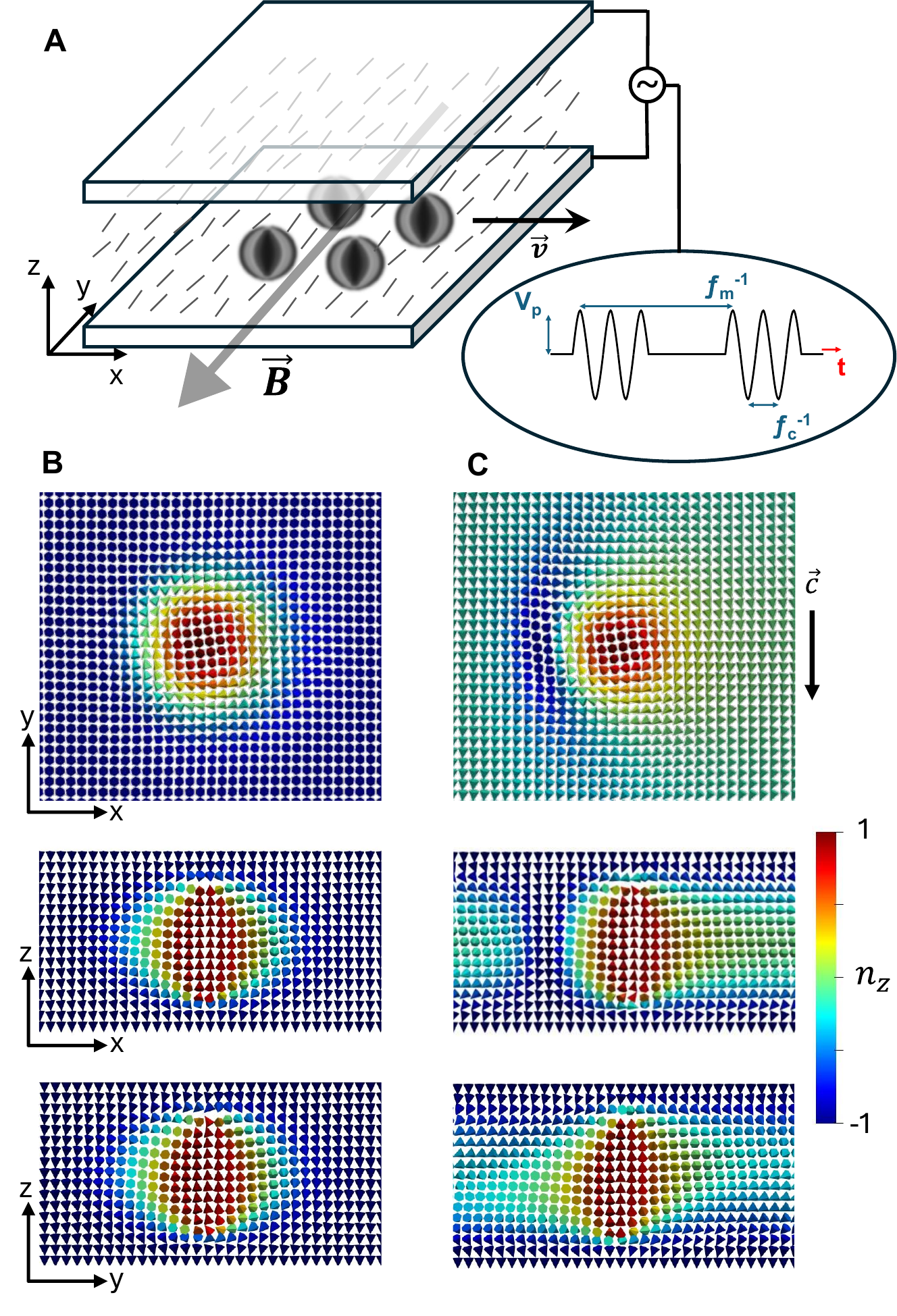}
\caption{(A) Sketch of the experimental setup. The liquid crystal fills the narrow gap between two ITO-coated glass plates, across which we apply an amplitude-modulated AC electric field. Torons of size comparable to the cell gap are formed within the liquid crystal layer. A magnetic field ($\bm{B}$) is applied parallel to the cell. (B) 3D structure of the director field, $\mathbf{n}$, within torons, in the absence of external fields, in the three orthogonal mid-planes. (C) Same as (B), but in the presence of an electric field between the plates. The vector $\vec{c}$ is the projection of $\mathbf{n}$ in the XY mid plane and away from the distortions. The configurations in (B) and (C) are obtained numerically after minimizing the free energy (see text).}
\label{Fig:setup}
\end{figure}

LC cells were made of ITO (Indium Tin Oxide)-coated glass slides (R=30$\Omega$-sq, purchased from Neyco, France) cut into 2.0 cm x 2.5 cm pieces. 
First, glasses were washed thoroughly with soap and water and, subsequently, with a 2\% water (Milli-Q)-surfactant (DECON®90, Merck) mixture in an ultrasound bath for 15 min. Next, glass pieces were rinsed with water (Milli-Q) in an ultrasound bath for 15 minutes. The cleaned glasses were dried under a stream of N$_2$ and their surface was activated using low-pressure oxygen plasma (Zepto M2, 0-50 W, purchased from Diener Electronic, Germany) for 180 s (oxygen pressure 0.4-0.6 mbar, power 45 W). 
Immediately afterwards, the surface was covered with a spin-coated (500 rpm for 30 s followed by 3 min at 3000 rpm) layer of a polyimide-based resin (3 Sunever 526, purchased from Nissan, Japan). Chemical bonding of the resin on the surface was ensured by baking the glasses at 180$^{\circ}$ C for 1 h, after a preheating process at 80$^{\circ}$ C for 5 min. This treatment ensures a strong homeotropic (perpendicular) anchoring of the LC on the surfaces.

Cells were assembled with the use of epoxy resin (Araldite purchased from Ceys, Spain) or photocurable glue (NOA81, purchased from Norland Products INC., USA). Tungsten wires (purchased from Goodfellow, UK) of different diameters were used as spacers. The electrical connection of the cell with the different devices was accomplished by attaching copper wires to the ITO substrates. Silver paint was applied between the copper wires and the cell to reinforce the connection. 

\subsection{Assembly of photopatterned cells for microfluidics} \label{Non-p-cell}

ITO-coated glass slides (R=30$\Omega$-sq purchased from Neyco, France) were cut into 1.0 cm x 2.5 cm pieces, washed, and surface-activated as indicated in section \ref{Non-p-cell}. They were subsequently spin-coated (500 rpm for 30 s followed by 3 min at 4000 rpm) with a layer of NOA81 photosensitive resin. 

Photopolymerized structures were created by using an inverted optical microscope modified with a digital micromirror device (DMD) UV light projector (Texas Instruments LightCrafter 4500, with a 2W 385nm LED, EKB Technologies, Ltd, Israel) to imprint the desired patterns. The latter were incorporated into the light path of the microscope with the use of a collimating lens (f= $+$140 mm) and a 505 nm dichoric mirror (Thorlabs DMLP505R, USA). The pattern was projected and focused onto the NOA81 plane by means of the microcope's 20x objective (UV light power density$=$ 3.1 $W·cm^{-2}$). Patterns were created using MS-PowerPoint software. The resultant solid structures had an approximate thickness of 6 to 9 $\mu$m after an irradiation time of 3 s. The glass was subsequently submerged in acetone to stop the polymerization reaction. An acetone bath with 30 min sonication was use to remove the uncured resin. Afterwards, we proceeded as explain in section \ref{sec:assembly}, starting from surface activation.  In this case, the channel structures acted also as spacers. 

\subsection{Cell filling}
Cells were filled with a mixture of the LC \textit{MBBA} (N-(4-methoxybenzylidene)-4-butylaniline) with the chiral LC \textit{CB15} ((S)-4-cyano-4'-[(2-methylbutyl)]biphenyl), both purchased from SYNTHON Chemicals, Germany. The mixtures were prepared by weight to obtain a certain pitch of the cholesteric twist ($P$) and, therefore, a target frustration value ($d/P$). $P$ as a function of the mass percentage of the chiral dopant was determined with the Grandjean-Cano wedge method \cite{OswaldBook}, measuring a helical twisting power $HTP = 7.9\pm 0.4\, \mu m^{-1} wt\%^{-1}$. 

\subsection{Steering and observation of torons}

Torons were generated using a low-frequency AC electric current to destabilize the LC, generating electrohydrodynamic instabilities (square-shaped wave with a peak amplitude ranging from 15 V to 20 V and a carrier frequency from 1 Hz to 50 Hz). Signals were created with a function generator (DSOX2002A InfiniiVision-Keysight, USA) and an electronic amplifier (WMA-100A-HV A-Falco Systems, Amsterdam). Depending on the amplitude and the frequency of the signal, the degree of destabilization could be controlled and, therefore, the areal density of torons was easily tuned within the cell. 
On the other hand, toron motion was triggered by an amplitude-modulated (AM) AC electric field with $100\,\%$ duty cycle. The sinusoidal-wave carrier frequency, $f_c$, varied from 1 to 6 kHz. The square-wave modulation frequency, $f_m$, was in the range 1 - 50 Hz (Fig. \ref{Fig:setup}). The permanent magnetic field of up to 0.4 T was created with a homemade Halback array of Nd magnets \cite{Ignes-Mullol20,Blumler2023PracticalResonance}. The assembly is placed over a mechanical stage that allowed the control of the magnetic field intensity and in-plane orientation.


Observation was performed with a custom-made polarized light optical microscope. Data acquisition was performed using a CMOS camera (PLA741-PixeLink, Canada). Samples were illuminated with a 625 nm led (ThorLabs M625L3, USA). 
Both the mechanical stage and optical microscope were described in detail in Ref. \cite{Ignes-Mullol20}.

Image analysis was performed with Image-J software. Single-toron tracking was performed with the software extension TrackMate in Image-J. The size of the torons was approximated as the area of a circle whose boundary was tangent to the outermost structure of the toron when the modulated AC electric field was applied.

\subsection{Simulations}

The liquid crystal is formed by elongated molecules which tend to align with their neighbours in the nematic phase. Their average orientation is described by the director field $\mathbf{n}(\mathbf{x})$~\cite{Doi2013}. The distortions in the director field are penalized by the elastic terms in the free energy. However, because the liquid crystal is cholesteric, it tends to form helicoidal structures with a cholesteric pitch $P$. The liquid crystal torons~\cite{PhysRevResearch.5.033210, Sohn2019} are modelled in our simulations using the Frank-Oseen free energy:

\begin{align}
\mathcal{F} &= \int d^3\mathbf{x} \biggl( 
  \frac{K_{11}}{2} (\nabla \cdot \mathbf{n})^2 
  + \frac{K_{22}}{2} \left[ \mathbf{n} \cdot (\nabla \times \mathbf{n}) + q_0 \right]^2 \notag \\
&\quad + \frac{K_{33}}{2} \left[ \mathbf{n} \times (\nabla \times \mathbf{n}) \right]^2  
  - K_{24} \left\{ \nabla \cdot \left[ \mathbf{n} (\nabla \cdot \mathbf{n}) \right. \right. \notag \\
&\quad \left. \left. + \mathbf{n} \times (\nabla \times \mathbf{n}) \right] \right\} 
  - \frac{\epsilon_0 \Delta \epsilon}{2} (\mathbf{E} \cdot \mathbf{n})^2 
  - \frac{\chi_a}{2\mu_0} (\mathbf{B} \cdot \mathbf{n})^2 \biggr),
\label{eq:free_energy}
\end{align}

\noindent where $K_{11}$, $K_{22}$, $K_{33}$, $K_{24}$ are the Frank elastic constants (bend, twist, splay and saddle-splay, respectively),  $\Delta \epsilon <0$ and $\chi_a>0$ are the electric and magnetic anisotropies and $q_0=2\pi/P$. We impose homeotropic (perpendicular) anchoring of the LC on all the boundaries. While we fix strong anchoring conditions on the top and bottom plates, we add an anchoring term to the free energy for the lateral walls: $f_w= -W_0(\mathbf{n}\cdot \mathbf{n}_w)^2/2$, where $W_0$ is the anchoring strength and $n_w$ is the preferential alignment direction, which we choose perpendicular to the walls in our simulations, according to the experimental realisation. The dynamics is governed by the relaxation of the free energy:
\begin{align}
\frac{\partial \mathbf{n} }{\partial t} = -\frac{1}{\gamma}  \frac{\delta \mathcal{F}}{\delta \mathbf{n}} ,
    \label{director-time-eq} 
\end{align}
where $\gamma$ is the rotational viscosity determining the rate of relaxation of the director field. This equation is solved using a predictor-corrector finite-differences algorithm.

The simulations are initialized with the directors pointing perpendicularly to the plates except in the proximity of a toron, whose configuration is obtained by minimizing the free energy starting from the {\it ansatz} of Ref.~\cite{Coelho_2021} and using the relaxation Eq. \ref{director-time-eq}. Next, we apply a constant electric field in the -z direction. To break the symmetry in the xy plane, we apply, during $0.1$ s, an oblique electric field with a component in the -y direction, making the total field become $\mathbf{E}=(0,\,-0.5,\, 0.5)E$. 
After this initial step, only the electric field along the -z direction is maintained, $\mathbf{E}=(0,\,0,\, 1)E$ being turned on and off with equal time intervals with a prescribed frequency, and a constant magnetic field is switched on. The electric field in the simulations has the same shape as in Fig.~\ref{Fig:setup}, but with $f_c=0$, i.e., it is a modulated DC field. Because there are no ionic impurities in the simulations,  there is no need for a high frequency carrier. The material parameters both in simulation and physical units, are given in Table~\ref{tab1}.

\begin{table}
\caption{\label{tab1} Parameters used in the model in simulation and physical units.}
\footnotesize
\begin{tabular}{|p{0.28\linewidth}|p{0.28\linewidth}|p{0.28\linewidth}|}
\hline
symbol&sim. units & physical units\\
\hline
$\rho$&1&1000 kg/m$^{3}$\\
\hline
$\Delta x$&1 & 0.8375 $\mu$m\\
\hline
$\Delta t$&1 & 1$\times10^{-6}$ s\\
\hline
$K_{11}$&$0.0129$& $6.37\times 10^{-12}$ N \\
\hline
$K_{22}$&$0.0061$& $3.0\times 10^{-12}$ N \\
\hline
$K_{33}$&$0.0142$& $7\times 10^{-12}$ N \\
\hline
$K_{24}$&$0.0015$& $0.75\times 10^{-12}$ N \\
\hline
$\gamma$& 111.3 & 0.078 Pa.s\\
\hline
$W_0$& 1.0 & 701.4 J/m$^3$\\
\hline
$\chi_a$& $10^{-6}$ & $10^{-6}$\\
\hline
$\Delta \epsilon$& -2 & -2\\
\hline
$P$ & 20 & 16.75 $\mu$m \\
\hline
$L_z$& 15 & 12.56 $\mu$m\\
\hline
\end{tabular}\\
\end{table}
\normalsize

\begin{figure}[t]
\begin{center}
\includegraphics[width=\columnwidth]{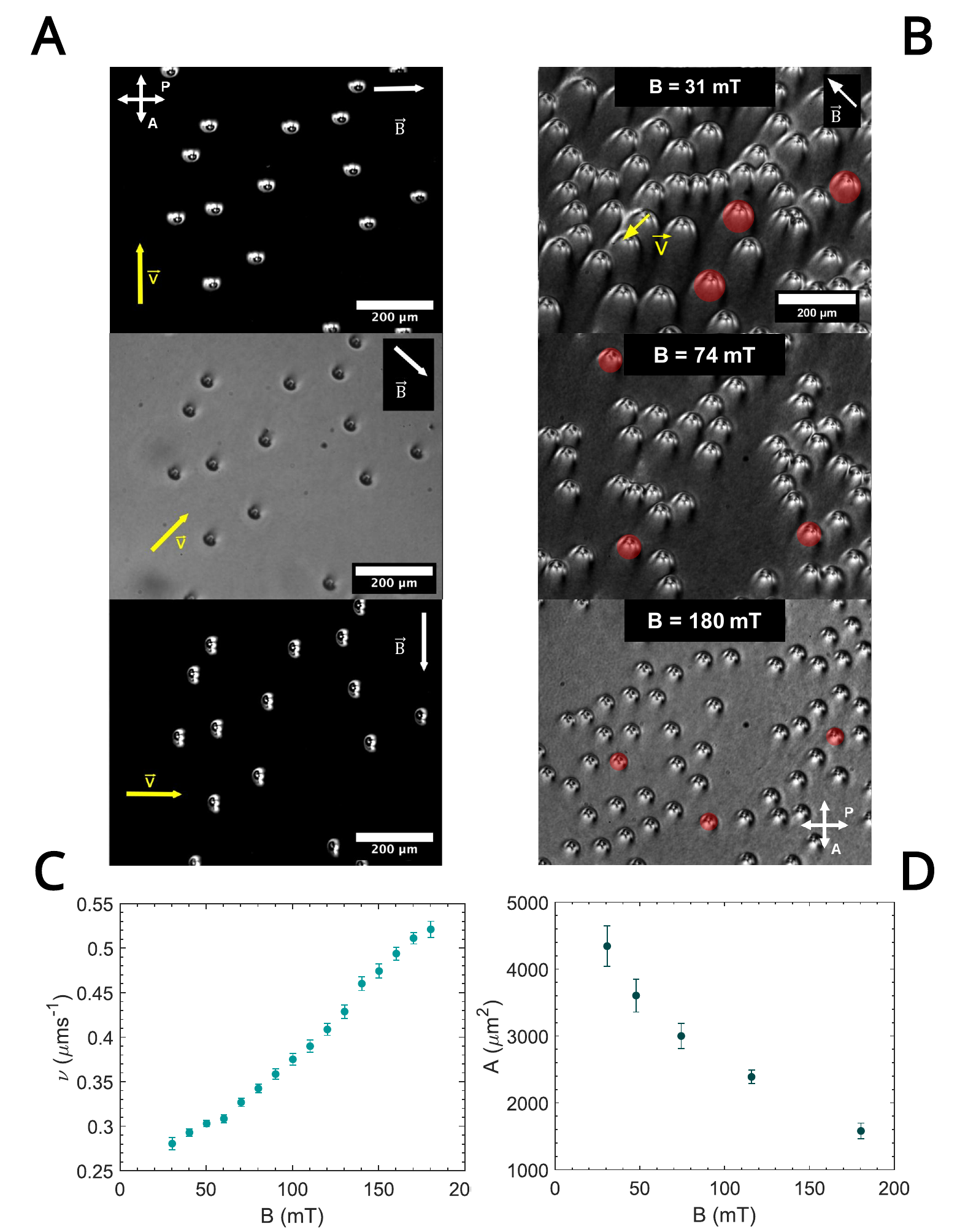}
\caption{(A) CLC cell ($d/P = 0.79$, $P = 23.5\,\mu$m) with torons being propelled by the action of an AC electric field with AM ($V_p=12$ V, $f_c=1$ kHz, $f_m=20$ Hz). The direction of motion (specified by the velocity vector) is selected through the application of an in-plane magnetic field ($B=400$ mT), as indicated in each panel. Observation  between crossed polarizers. See also Movie \ref{SMov:skyrmion_H_exp}A. (B) Effect of magnetic field intensity on toron size. Field and velocity directions are shown in the top panel. Observation between crossed polarizers. See also Movie \ref{SMov:skyrmion_H_exp}B. Speed (C) and size (D) of torons as a function of the magnetic field intensity, for the same AC driving. The red semi-transparent circles in panel B illustrate the area measurements shown in panel D, which were determined as described in Section \textit{2.4}.}
\label{Fig:steering}
\end{center}
\end{figure}

\begin{figure}[t]
\includegraphics[width=\columnwidth]{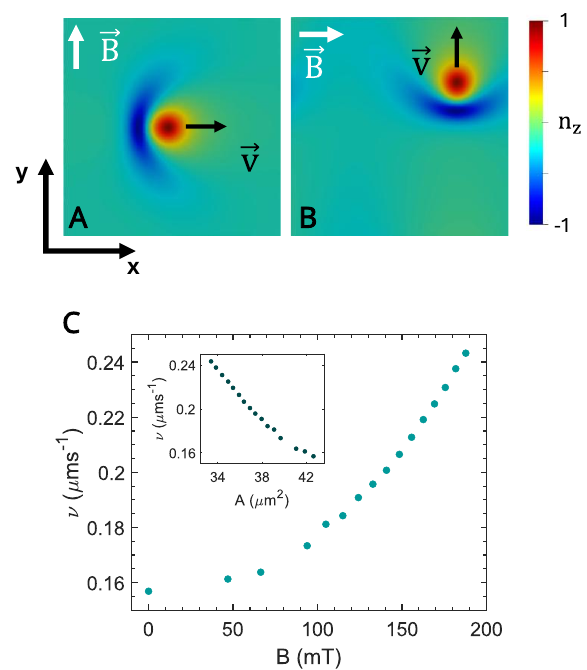}
\caption{(A-B) Simulations of a toron being driven by an amplitude-modulated DC electric field and oriented with a magnetic field. Color map corresponds to the out-of-plane component of the local director field (see also Movie \ref{SMov:skyrmion_H_sim}). The orientation of $\bm{B}$ is changed 90 degrees clockwise between panels A and B. 
(C) Toron speed as a function of the magnetic field amplitude. In the inset, and for the same data, the toron speed is plotted vs the area (computed with the condition n$_{z}$\,$>0$) of its intersection with the XY plane, obtained with the condition $n_z > 0$. Periodic boundary conditions are considered in both directions in a system of size $107\,\mu$m $\times$ $107\,\mu$m. Cell parameters are $d/P = 0.79$, $P = 23.5\,\mu$m. Electric field parameters are $V_p=3.97$ V, $f_m=20$ Hz.  A magnetic field of 0.19 T is applied as indicated in A and B.}
\label{Fig:steering_sim}
\end{figure}

\section{Results and discussion}

Due to their tendency to align at a small but finite angle with their neighbors, molecules in a cholesteric liquid crystal (CLC) organize spontaneous twist distortions of the director field, characterized by its periodicity or pitch, $P$ \cite{OswaldBook,Kleman2003}. This ground state configuration can be frustrated by confining a layer of CLC thinner than $P$ between two plates that impose homeotropic (perpendicular) boundary conditions on their surface. When the cell gap, $d$, is larger than $0.7 P$, approximately, a variety of solitonic topological structures can emerge \cite{NAwa1978,Wu2022}. In particular, strong homeotropic anchoring conditions lead to the formation of three-dimensional localized structures called torons, often characterized by their midplane cuts named skyrmions because of their topological analogy with the vortex-like spin structures that emerge in certain magnetic materials \cite{Foster2019Two-dimensionalFerromagnets,Sohn2019,deSouza2025}. 
Fig. \ref{Fig:setup}B shows the expected three dimensional structure of an equilibrium LC toron, obtained my minimizing the elastic free energy (Eq. \ref{eq:free_energy}). Its vertical cross sections reveal that torons are limited by a topological defects near each of the bounding plates \cite{Leonov2014,Tai2020}. Their size in all direction is of the order of $d$ and, thus, $P$. We generate torons by applying a large amplitude, low frequency, AC electric field across the LC layer for a few seconds. The presence of ionic impurities in the CLC sample triggers electrohydrodynamic instabilities and chaotic convection patterns. Suddenly switching off the field leads to LC relaxation back into the homeotropic nematic configuration and to the spontaneous formation of stable torons \cite{Ackerman2017}. For the used LC mixtures and cell gap values, this is typically achieved with AC electric fields with a carrier frequency $f_c\,=10$ Hz and amplitude $V_{p}\approx 20$ V. Alternatively, torons can be generated by the slow transition from the isotropic to the nematic phase or, more locally, by using tightly focused laser beams. 

In the absence of external fields and at constant temperature, torons remain stable, in spite of their accumulated elastic energy. Application of a strong enough electric field perpendicular to the two transparent electrodes leads to the reorientation of $\mathbf{n}$ towards the plane of the cell (translationally-invariant configuration, or TIC \cite{OswaldBook}), a result of the negative dielectric anisotropy of the mesogen MBBA (see Fig. \ref{Fig:setup}C for the 3D structure of a toron under electric field). The average in-plane direction defines a $\vec{c}$-director, whose orientation is normally random in the absence of additional influences that break the symmetry \cite{Oswald2023}. While transition to TIC under an electric field results in the annihilation of existing torons, it has been recently shown that application of an AM AC electric field can distort the torons structure without destroying them, breaking their fore-aft symmetry and leading to their net propulsion as solitonic waves in a direction perpendicular to $\vec{c}$ (Fig. \ref{Fig:setup}C) \cite{Ackerman2017,Sohn2018}. This behavior is similar to that of phoretic colloidal inclusions embedded in LC's \cite{Lavrentovich2016,Peng2019}. Different from their solid counterparts, however, toron quasiparticles are intrinsically identical, can be created or annihilated at will, and do not require surface treatments that cause heterogeneities in phoretic particles. 

Being able to set and reconfigure the direction of propulsion of LC torons is crucial for their applicability in microfluidic environments. The $\vec{c}$-director can be established by gentle unidirectional rubbing of the anchoring layer upon sample preparation\cite{Ackerman2017}, but this procedure lacks reconfigurability. It can also be locally set using linearly-polarized light if a suitable chromophore is dissolved with the mesogen \cite{Sohn2019b,Sohn2020}. Instead, here we demonstrate that full toron steering can be achieved by means of a magnetic field, thus not needing additional impurities to be added within the CLC mixture. 

In Fig. \ref{Fig:steering}A, we demonstrate the ability to steer driven torons with the homogeneous, in-plane magnetic field created by an array of permanent magnets, for different orientations of $\vec{B}$. For the ratio $d/P$ used here, $\vec{c}$ orients parallel to $\vec{B}$. Hence, the colloidal quasiparticles are propelled in a direction perpendicular to $\vec{B}$. Interestingly, propulsion parallel to $\vec{B}$ can also be achieved for $d/P \gtrsim 0.95$ as, in that case, $\vec{c}$ orients perpendicular to $\vec{B}$ \cite{Oswald2023}. An example of the robustness of the driven torons is obtained with a wedge cell featuring a spatial gradient in $d$. We can observe how torons turn 90 degrees when the boundary $d/P \simeq 0.95$ is crossed, while retaining their integrity (See Movie \ref{SMov:twoTICs}). 

Torons shrink and move faster when the amplitude of $\vec{B}$ increases (Fig. \ref{Fig:steering}A-D). In fact, a similar correlation between toron size and speed is observed when they are studied at different amplitudes of the modulated AC electric field, as torons are smaller and move faster when the AC electric field amplitude is increased  (Fig. \ref{SFig:EField}A,B). Moving torons are propagating solitary waves, rather than associated with a net material flow \cite{Leonov2014,Sohn2018,Alvim2024}. Their motion implies the local reorientation of the mesogen molecules, which are affected by the material's rotational viscosity. As the region influenced by these reorientations becomes smaller when torons shrink, it is reasonable that their propagation speed increases. The latter also increases steadily with the modulation frequency of the AC electric field, reaching a maximum at around 40 Hz (Fig. \ref{SFig:EField}C).

The driving and steering of LC torons reported above has been also studied in numerical simulations, as detailed in the Methods section. In Fig. \ref{Fig:steering_sim}A,B we include the evolution of the cross section along the XY plane of a LC toron (skyrmion) when driven by an AM electric field (oriented along the Z axis).
In this regime, $\bm{c}$ is parallel to $\bm{B}$ and the torons, which always move perpendicularly to $\vec{c}$, also move perpendicularly to $\bm{B}$. Between panels A and B in Fig. \ref{Fig:steering_sim}, the magnetic field is suddenly rotated 90 degrees clockwise. As a consequence, $\bm{c}$ slowly reorients to be parallel to  $\bm{B}$, which triggers a similar change in the toron's trajectory. Simulations reproduce the speed increase with the magnetic field amplitude and the concomitant shrinkage of the toron's cross-sectional area. In simulations, this area is defined by the condition $n_z > 0.1$ (see Fig. \ref{Fig:steering_sim}C).

\begin{figure}[t]
\begin{center}
\includegraphics[width=0.8\columnwidth]{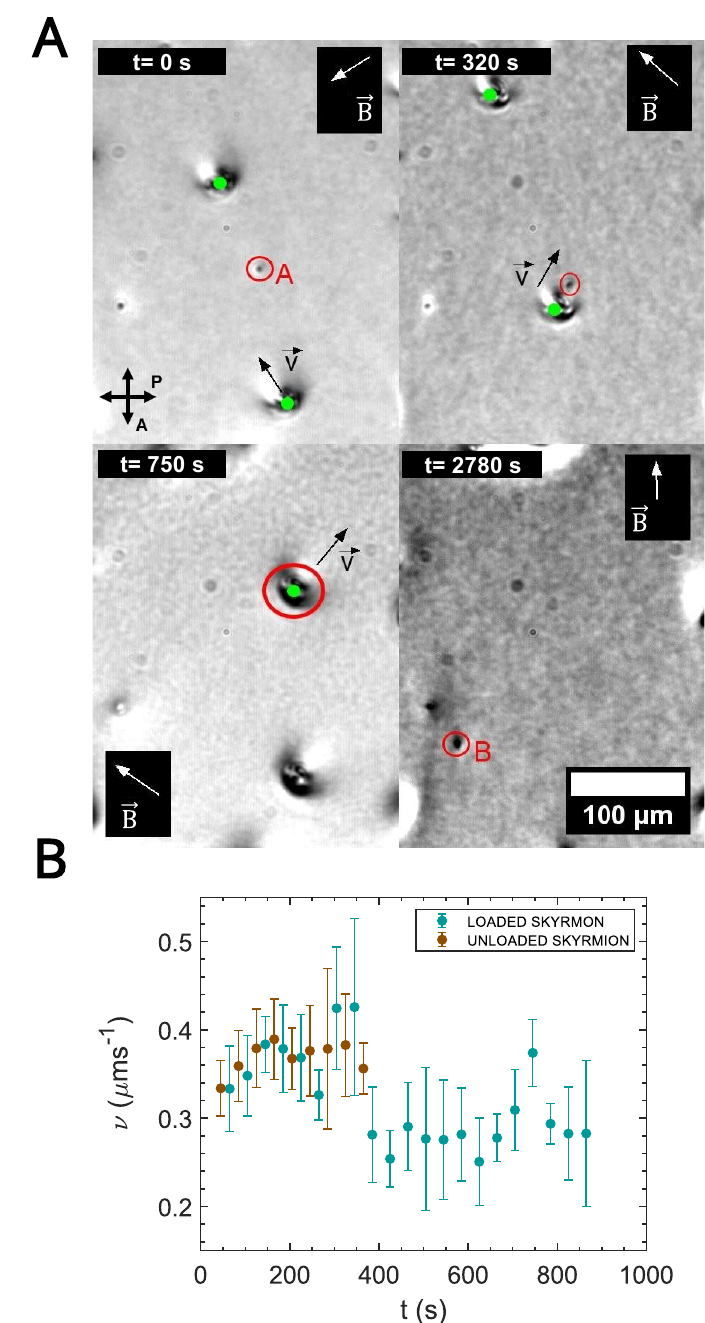}
\end{center}
\caption{
Controlled transport of a passive polystyrene particle with planar LC anchoring on its surface by a toron ($d/P$=0.77, $p$=23.5 $\mu$m) driven by the action of an AM AC electric field ($V_p$=12.7 V, $f_m=20$ Hz $f_c=1$ kHz) and steered with an in-plane permanent magnetic field ($B = 0.4$ mT). (A) POM images of the particle transport. The red circle marks the location of the particle. The green dot marks the location of torons in the field of view. See also Movie \ref{SMov:transport}. Elapsed times are indicated. (B) Instantaneous speed of the toron that is loaded at 350 s of elapsed time. Error bars refer to the confidence interval of each measurement.
}
\label{Fig:transport}
\end{figure}

As discussed above, moving torons are solitary waves without mass transport. However, we can take advantage of the tendency colloidal inclusions have to bind to LC spatial distortions to use driven torons as cargo transport vectors \cite{Sohn2018} . In Fig. \ref{Fig:transport}A we present the protocol to achieve this in a controlled manner by taking advantage of the application of an external magnetic field. A driven toron is directed towards a passive colloidal particle by suitable orientation of the magnetic field. The particle is absorbed by the toron at 330 s of elapsed time. The ensemble of toron and colloidal cargo is then driven along an arbitrary path with the usual toron transport protocol until reaching a target final destination. At this point, torons are annihilated by suddenly increasing the electric field amplitude ($V_p$= 28 V in this experiment). This releases the colloidal cargo in the new location. As shown in Fig. \ref{Fig:transport}B, mobility decreased for loaded torons, but they can still be driven with the usual mechanism. We can enhance the versatility of this mass transport protocol by the \textit{ad-hoc} creation of torons on top of cargo-carrying particles. In Movie \ref{SMov:laser} we use a laser source incorporated in the light path of the microscope to dress a passive colloidal particle with a toron, which is later driven and eventually destroyed in a new location, thus delivering the cargo. 

\begin{figure}[t]
\begin{center}
\includegraphics[width=\columnwidth]{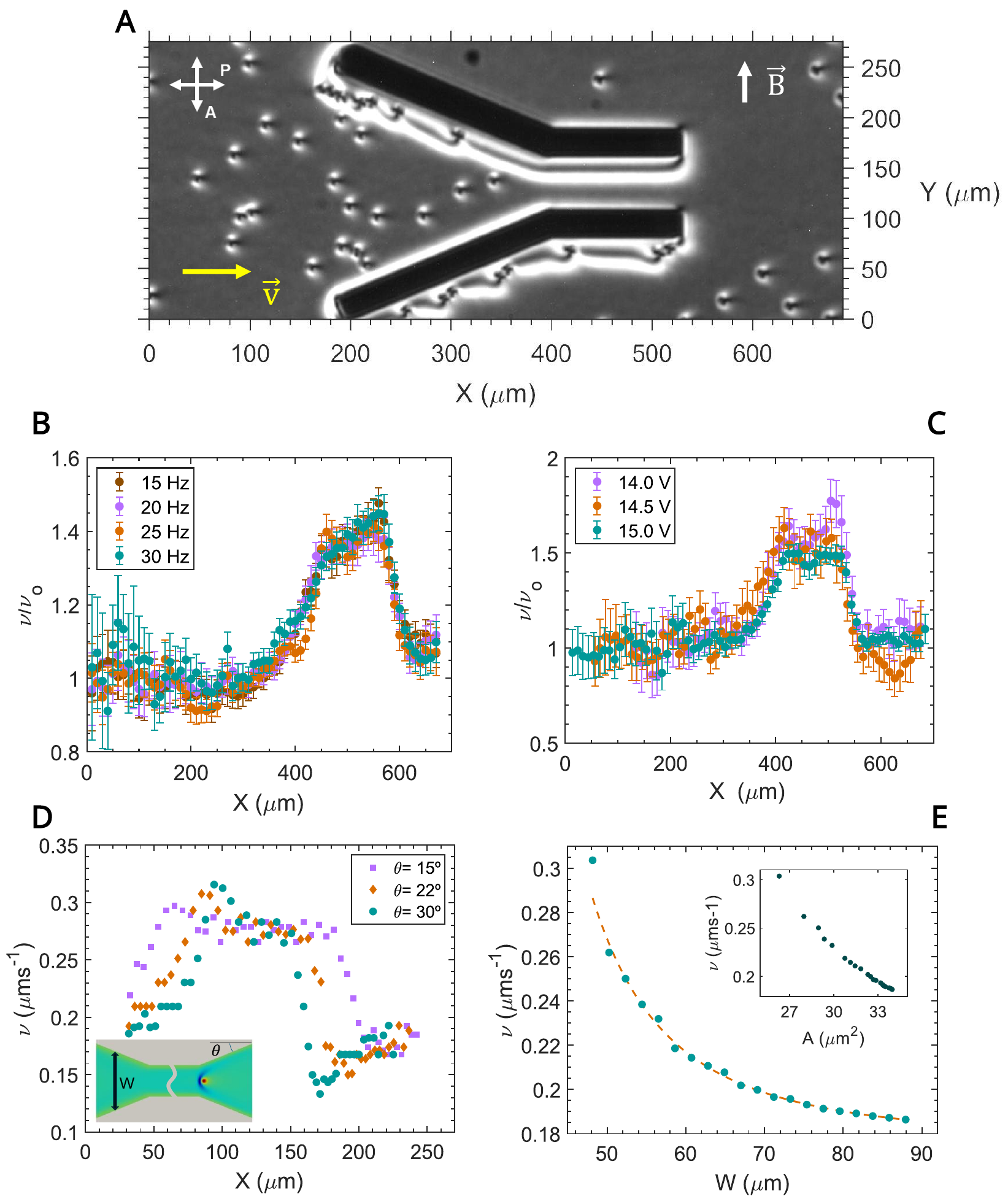}
\end{center}
\caption{(A) Ensemble of torons driven through a microfluidic funnel into a microchannel (see also Movie \ref{SMov:funnel}). Cell parameters are $d/P=0.7$, $P=11.6\,\mu$m. (B) Average speed profile along the traveling direction for different $f_m$ ($V_p$ =14 V, $f_c$ = 2 kHz). (C) Average speed profile along the traveling directions for different $V_p$ ($f_m$ = 20 Hz, $f_c$ = 2 kHz). All speeds are normalized with the average value outside the microfluidic device. Error bars refer to the confidence interval of each measurement. (D) Simulations of torons driven along a double-funnel microfluidic device for different angular openings, $\theta$ (see also Movie \ref{SMov:sim_funnel}). The length of the straight part decreases with $\theta$.  System size is $251\,\mu$m $\times$ $168\,\mu$m, with strong boundary conditions on the walls, and periodic boundary conditions in the horizontal direction. Cell parameters are $d/P=0.75$, $P=16.75\,\mu$m. AC modulation conditions are $V_p$ =3.75 V, $f_m$ = 20 Hz. A magnetic field of 0.12 T is applied in the Y direction.  (E) Toron speed vs. channel width from simulations within a uniform channel of width $W$. In the inset, speed vs. toron area for the same simulations. Same conditions as in (D). 
}
\label{Fig:YjunctionExp}
\end{figure}

The behavior of driven torons as colloidal quasiparticles is finally addressed by directing an ensemble of them through a microfluidic device embedded within the LC cell (see Materials and Methods). We built microstructures in the form of funnels that focus the torons through a narrow passage of width less than twice their size. In Fig. \ref{Fig:YjunctionExp}A, the magnetic field is oriented so that toron velocities are parallel to the device's length. The LC has homeotropic boundary conditions both on the top and bottom plates and on the device's walls. As a result of the LC distortion, torons are elastically repelled by the walls. We observe, however, some torons trapped by disclination lines that run parallel to the South-facing walls of the device, likely a consequence of asymmetries in the spin-coating process during device preparation to render all walls homeotropic (see Sec. \ref{Non-p-cell}). Torons entering the funnel self-organize into a single row before crossing the channel in sequence. In other words, this device transforms the initially disordered swarm into a jet of individual quasiparticles. Interestingly, toron speed decreases steadily with the local wall separation, which depends on the position along the channel in the funnel geometry. The lowest speed is thus observed outside the device while the maximum one is found within the narrow passage. This geometrical modulation of the propulsion speed relative to that outside the microfluidic device remains the same after changes in the AM modulation parameters (Fig. \ref{Fig:YjunctionExp}B,C). Also, this dependence on the local channel width does not seem to be significantly influenced by geometrical details of the device, such as the aperture angle of the funnel (Fig. \ref{SFig:Yjunction_exp}).  

Numerical simulations of a double-funnel channel confirm the experimental observations (Fig. \ref{Fig:YjunctionExp}D), including the independence of speed modulations on features of the channel geometry, such as its angular opening. A detailed analysis of the dependence of toron speed and size on channel width (Fig. \ref{Fig:YjunctionExp}D) shows again a correlation between those two parameters. For the conditions of these simulations, torons begin to shrink when wall spacing falls below $90\,\mu$m, approximately, with a concomitant increase in their speed. Lateral confinement with repulsive walls acts as an external field that progressively compresses torons to a smaller size, thus increasing their mobility. Inspection of the simulated data suggests that the speed follows an exponential decay with channel width, defining a penetration depth of the wall repulsion of $11.2\pm1.2\,\mu$m, similar to $P$ in these simulations. The wall effect strongly depends on the anchoring strength imposed on the walls, vanishing for weak anchoring where no distortion is imposed by the walls into the bulk LC (Fig. \ref{SFig:anchoring}A). 

Simulations also suggest that not only the channel width but also the distance to the wall has a weak influence on toron speed, with higher values closer to the walls (Fig. \ref{SFig:anchoring}B; see also Movie \ref{SMov:column}). Experiments of torons moving at different distances to a wall are consistent with these findings (Fig. \ref{SFig:Expvelstraightwall}; see also Movie \ref{SMov:wall}). This result, however, does not seem to affect the dynamics in the funnel-shaped devices. Indeed, most torons entering a device are eventually deflected by a lateral wall, since walls in the funnler devices are not parallel to the direction of motion. Interestingly, torons continue travelling parallel to the wall but with the same speed as those moving near the center of the device (Fig. \ref{SFig:Yjunction_exp_comparison}; see also Movie \ref{SMov:funnel}). Notice that torons traveling parallel to the wall are repelled by it, keeping a distance of around $30\,\mu$m, at which, as discussed above, speed is barely affected by the presence of the wall. This behavior is maintained as long as the wall orientation remains within $50\,^{\circ}$ of the original direction of toron motion. If torons collide with the wall at a higher angle, they are absorbed by the LC distortion at the boundary. This result is important in order to apply our proposed steering mechanism with complex microfluidic channels. Indeed, all torons within a sample are steered along the same direction by the uniform magnetic field. However, an embedded device featuring walls with different orientations will still benefit from our strategy, as torons will continue to travel along all walls with the same propulsion speed, provided their orientation is within the above exposed limits.

\section{Conclusions}


In this work, we have demonstrated that cholesteric liquid crystal torons, a type of topological quasiparticle, can be created, propelled, and magnetically steered within confined liquid crystal layers. Combining experiments and numerical simulations, we showed that these solitonic excitations, despite carrying no net mass flow, can function as addressable micro-carriers, capable of capturing, transporting, and releasing colloidal cargo under precise external control using combined electric and magnetic fields. We further integrated torons into microfluidic devices, uncovering how lateral confinement enhances their stability and mobility due to elastic interactions with channel walls. This establishes torons as a robust tool for programmable transport in soft matter systems, with advantages over conventional active colloids, as they are intrinsically uniform, reconfigurable, and do not rely on surface functionalization.


Beyond their applicability in microfluidics, our findings highlight liquid crystal torons as a versatile platform at the interface of soft matter physics, materials science, and microrobotics. Their soft and deformable nature, combined with topological protection, provides a unique opportunity to explore emergent collective phenomena such as clogging, self-organization, and active jamming in colloidal flows. At the same time, their ability to be locally created and annihilated paves the way for reconfigurable metamaterials and adaptive transport systems that respond dynamically to external stimuli. Taken together, our results show that torons are not only an ideal model system for active colloidal matter, but also a building block for next-generation soft robots and adaptive lab-on-chip devices, bridging fundamental physics with technological applications.

\section*{Acknowledgements}
J.I.-M., F.S. and J.T.-A. acknowledge funding from MICIU/AEI/10.13039/501100011033 (Grant No. PID2022-137713NB-C21). J.T.-A. acknowledges the support and funding of a FPU fellowship FPU22/01916 from MICIU. R. C. acknowledges financial support from the Portuguese Foundation for Science and Technology (FCT) under the contracts: UIDB/00618/2020 (DOI 10.54499/UIDB/00618/2020), UIDP/00618/2020 (DOI 10.54499/UIDP/00618/2020), and 2023.10412.CPCA.A2 (DOI 10.54499/2023.10412.CPCA.A2). J.I.-M. acknowledges the financial support and hospitality from ENS de Lyon.

\bibliographystyle{elsarticle-num} 
\bibliography{cas-refs}

\begin{thebibliography}{10}
\expandafter\ifx\csname url\endcsname\relax
  \def\url#1{\texttt{#1}}\fi
\expandafter\ifx\csname urlprefix\endcsname\relax\def\urlprefix{URL }\fi
\expandafter\ifx\csname href\endcsname\relax
  \def\href#1#2{#2} \def\path#1{#1}\fi

\bibitem{Bishop2023}
K.~J.~M. Bishop, S.~L. Biswal, B.~Bharti, Active colloids as models, materials,
  and machines, Annu Rev Chem Biomol Eng 14 (2023) 1--30.

\bibitem{Yan2022}
G.~Yan, A.~A. Solovev, G.~Huang, J.~Cui, Y.~Mei, Soft microswimmers: Material
  capabilities and biomedical applications, Current Opinion in Colloid and
  Interface Science 61 (10 2022).
\newblock \href {https://doi.org/10.1016/j.cocis.2022.101609}
  {\path{doi:10.1016/j.cocis.2022.101609}}.

\bibitem{Bunea2020}
A.~I. Bunea, R.~Taboryski, Recent advances in microswimmers for biomedical
  applications, Micromachines 11 (2020) 1--24.
\newblock \href {https://doi.org/10.3390/mi11121048}
  {\path{doi:10.3390/mi11121048}}.

\bibitem{Sharan2021}
P.~Sharan, A.~Nsamela, S.~C. Lesher-Pérez, J.~Simmchen, Microfluidics for
  microswimmers: Engineering novel swimmers and constructing swimming lanes on
  the microscale, a tutorial review, Small 17 (7 2021).
\newblock \href {https://doi.org/10.1002/smll.202007403}
  {\path{doi:10.1002/smll.202007403}}.

\bibitem{Howse2007}
J.~R. Howse, R.~A.~L. Jones, A.~J. Ryan, T.~Gough, R.~Vafabakhsh,
  R.~Golestanian, Self-motile colloidal particles: From directed propulsion to
  random walk, Phys. Rev. Lett. 99 (2007) 048102.
\newblock \href {https://doi.org/10.1103/PhysRevLett.99.048102}
  {\path{doi:10.1103/PhysRevLett.99.048102}}.

\bibitem{Hong2007}
Y.~Hong, N.~M.~K. Blackman, N.~D. Kopp, A.~Sen, D.~Velegol, Chemotaxis of
  nonbiological colloidal rods, Phys. Rev. Lett. 99 (2007) 178103.
\newblock \href {https://doi.org/10.1103/PhysRevLett.99.178103}
  {\path{doi:10.1103/PhysRevLett.99.178103}}.

\bibitem{Bricard2013}
A.~Bricard, J.~B. Caussin, N.~Desreumaux, O.~Dauchot, D.~Bartolo, Emergence of
  macroscopic directed motion in populations of motile colloids, Nature 503
  (2013) 95--98.
\newblock \href {https://doi.org/10.1038/nature12673}
  {\path{doi:10.1038/nature12673}}.

\bibitem{Gangwal2008}
S.~Gangwal, O.~J. Cayre, M.~Z. Bazant, O.~D. Velev, Induced-charge
  electrophoresis of metallodielectric particles, Phys. Rev. Lett. 100 (2008)
  058302.
\newblock \href {https://doi.org/10.1103/PhysRevLett.100.058302}
  {\path{doi:10.1103/PhysRevLett.100.058302}}.

\bibitem{Tierno2007}
P.~Tierno, T.~H. Johansen, T.~M. Fischer, Magnetically driven colloidal
  microstirrer, Journal of Physical Chemistry B 111 (2007) 3077--3080.
\newblock \href {https://doi.org/10.1021/jp070579o}
  {\path{doi:10.1021/jp070579o}}.

\bibitem{Liao2019}
P.~Liao, L.~Xing, S.~Zhang, D.~Sun, Magnetically driven undulatory
  microswimmers integrating multiple rigid segments, Small 15 (9 2019).
\newblock \href {https://doi.org/10.1002/smll.201901197}
  {\path{doi:10.1002/smll.201901197}}.

\bibitem{Lozano2016}
C.~Lozano, B.~T. Hagen, H.~Löwen, C.~Bechinger, Phototaxis of synthetic
  microswimmers in optical landscapes, Nature Communications 7 (9 2016).
\newblock \href {https://doi.org/10.1038/ncomms12828}
  {\path{doi:10.1038/ncomms12828}}.

\bibitem{Arya2020}
P.~Arya, D.~Feldmann, A.~Kopyshev, N.~Lomadze, S.~Santer, Light driven guided
  and self-organized motion of mesoporous colloidal particles, Soft Matter 16
  (2020) 1148--1155.
\newblock \href {https://doi.org/10.1039/c9sm02068c}
  {\path{doi:10.1039/c9sm02068c}}.

\bibitem{Piazza2008}
R.~Piazza, A.~Parola, Thermophoresis in colloidal suspensions, Journal of
  Physics Condensed Matter 20 (4 2008).
\newblock \href {https://doi.org/10.1088/0953-8984/20/15/153102}
  {\path{doi:10.1088/0953-8984/20/15/153102}}.

\bibitem{kolacz2020}
J.~Kołacz, A.~Konya, R.~L.~B. Selinger, Q.-H. Wei, Thermophoresis of colloids
  in nematic liquid crystal, Soft Matter 16 (2020) 1989--1995.
\newblock \href {https://doi.org/10.1039/C9SM02424G}
  {\path{doi:10.1039/C9SM02424G}}.

\bibitem{Katuri2017}
J.~Katuri, X.~Ma, M.~M. Stanton, S.~Sánchez, Designing micro-and nanoswimmers
  for specific applications, Accounts of Chemical Research 50 (2017) 2--11.
\newblock \href {https://doi.org/10.1021/acs.accounts.6b00386}
  {\path{doi:10.1021/acs.accounts.6b00386}}.

\bibitem{XU2022121551}
Y.~Xu, Q.~Bian, R.~Wang, J.~Gao, Micro/nanorobots for precise drug delivery via
  targeted transport and triggered release: A review, International Journal of
  Pharmaceutics 616 (2022) 121551.
\newblock \href {https://doi.org/https://doi.org/10.1016/j.ijpharm.2022.121551}
  {\path{doi:https://doi.org/10.1016/j.ijpharm.2022.121551}}.

\bibitem{WANG2013531}
W.~Wang, W.~Duan, S.~Ahmed, T.~E. Mallouk, A.~Sen, Small power: Autonomous
  nano- and micromotors propelled by self-generated gradients, Nano Today 8~(5)
  (2013) 531--554.
\newblock \href {https://doi.org/https://doi.org/10.1016/j.nantod.2013.08.009}
  {\path{doi:https://doi.org/10.1016/j.nantod.2013.08.009}}.

\bibitem{Ju2025}
X.~Ju, Chen, et~al., Technology roadmap of micro/nanorobots, ACS Nano (2025).

\bibitem{Doostmohammadi2018}
A.~Doostmohammadi, J.~Ignés-Mullol, J.~M. Yeomans, F.~Sagués, Active nematics
  (12 2018).
\newblock \href {https://doi.org/10.1038/s41467-018-05666-8}
  {\path{doi:10.1038/s41467-018-05666-8}}.

\bibitem{Fu2017}
S.~Fu, F.~Wei, C.~Yin, L.~Yao, Y.~Wang, {Biomimetic soft micro-swimmers: from
  actuation mechanisms to applications} (2017).
\newblock \href {https://doi.org/10.1007/s10544-021-00546-3/Published}
  {\path{doi:10.1007/s10544-021-00546-3/Published}}.

\bibitem{C0SM00953A}
E.~Lauga, Life around the scallop theorem, Soft Matter 7 (2011) 3060--3065.
\newblock \href {https://doi.org/10.1039/C0SM00953A}
  {\path{doi:10.1039/C0SM00953A}}.

\bibitem{Yeomans2014}
J.~M. Yeomans, D.~O. Pushkin, H.~Shum, An introduction to the hydrodynamics of
  swimming microorganisms, European Physical Journal: Special Topics 223 (2014)
  1771--1785.
\newblock \href {https://doi.org/10.1140/epjst/e2014-02225-8}
  {\path{doi:10.1140/epjst/e2014-02225-8}}.

\bibitem{Ignes-Mullol2022}
J.~Ign\'es-Mullol, F.~Sagu\'s, Experiments with active and driven synthetic
  colloids in complex fluids, Current Opinion in Colloid \& Interface Science 62
  (2022).

\bibitem{Smalyukh2025LiquidColloids}
I.~I. Smalyukh, {Liquid Crystal Colloids} 50 (2025) 8.
\newblock \href {https://doi.org/10.1146/annurev-conmatphys}
  {\path{doi:10.1146/annurev-conmatphys}}.

\bibitem{Peng2019}
C.~Peng, O.~D. Lavrentovich, Liquid crystals-enabled ac electrokinetics (1
  2019).
\newblock \href {https://doi.org/10.3390/mi10010045}
  {\path{doi:10.3390/mi10010045}}.

\bibitem{Lavrentovich2016}
O.~D. Lavrentovich, Active colloids in liquid crystals, Current Opinion in
  Colloid \& Interface Science 21 (2016) 97--109.

\bibitem{Fumeron2023}
S.~Fumeron, B.~Berche, Introduction to topological defects: from liquid
  crystals to particle physics, European Physical Journal: Special Topics 232
  (2023) 1813--1833.
\newblock \href {https://doi.org/10.1140/epjs/s11734-023-00803-x}
  {\path{doi:10.1140/epjs/s11734-023-00803-x}}.

\bibitem{OswaldBook}
P.~Oswald, P.~Pieranski, Nematic and cholesteric liquid crystals : concepts and
  physical properties illustrated by experiments, Taylor \& Francis, Boca
  Raton, 2005.

\bibitem{Doi2013}
M.~Doi, Soft Matter Physics, Oxford University Press, 2013.
\newblock \href {https://doi.org/10.1093/acprof:oso/9780199652952.001.0001}
  {\path{doi:10.1093/acprof:oso/9780199652952.001.0001}}.

\bibitem{Kleman2003}
M.~Kleman, O.~D. Lavrentovich, Soft Matter Physics: An Introduction, 1st
  Edition, Springer-Verlag New York, Inc, 2003.

\bibitem{Dierking2005AnnihilationFields}
I.~Dierking, O.~Marshall, J.~Wright, N.~Bulleid, {Annihilation dynamics of
  umbilical defects in nematic liquid crystals under applied electric fields},
  Physical Review E - Statistical, Nonlinear, and Soft Matter Physics 71~(6) (6
  2005).
\newblock \href {https://doi.org/10.1103/PhysRevE.71.061709}
  {\path{doi:10.1103/PhysRevE.71.061709}}.

\bibitem{Wu2022}
J.-S. Wu, I.~I. Smalyukh, Hopfions, heliknotons, skyrmions, torons and both
  abelian and nonabelian vortices in chiral liquid crystals, Liquid Crystals
  Reviews (2022) 1--35.

\bibitem{Sohn2018}
H.~R. Sohn, P.~J. Ackerman, T.~J. Boyle, G.~H. Sheetah, B.~Fornberg, I.~I.
  Smalyukh, Dynamics of topological solitons, knotted streamlines, and
  transport of cargo in liquid crystals, Physical Review E 97 (5 2018).
\newblock \href {https://doi.org/10.1103/PhysRevE.97.052701}
  {\path{doi:10.1103/PhysRevE.97.052701}}.

\bibitem{Ackerman2017}
P.~J. Ackerman, T.~Boyle, I.~Smalyukh, Squirming motion of baby skyrmions in
  nematic fluids, Nat Commun 8~(1) (2017) 673.

\bibitem{Shen2022RecentCrystals}
Y.~Shen, I.~Dierking, {Recent Progresses on Experimental Investigations of
  Topological and Dissipative Solitons in Liquid Crystals} (1 2022).
\newblock \href {https://doi.org/10.3390/cryst12010094}
  {\path{doi:10.3390/cryst12010094}}.

\bibitem{Sohn2019}
H.~R.~O. Sohn, C.~D. Liu, I.~I. Smalyukh, Schools of skyrmions with
  electrically tunable elastic interactions, Nature Communications 10~(1)
  (2019).
\newblock \href {https://doi.org/10.1038/s41467-019-12723-3}
  {\path{doi:10.1038/s41467-019-12723-3}}.

\bibitem{Hernandez-Navarro2014}
S.~Hernandez-Navarro, P.~Tierno, J.~A. Farrera, J.~Ignes-Mullol, F.~Sagues,
  Reconfigurable swarms of nematic colloids controlled by photoactivated
  surface patterns, Angew Chem Int Ed Engl 53~(40) (2014) 10696--700.

\bibitem{Tai2020}
J.~S.~B. Tai, I.~I. Smalyukh, Surface anchoring as a control parameter for
  stabilizing torons, skyrmions, twisted walls, fingers, and their hybrids in
  chiral nematics, Physical Review E 101 (4 2020).
\newblock \href {https://doi.org/10.1103/PhysRevE.101.042702}
  {\path{doi:10.1103/PhysRevE.101.042702}}.

\bibitem{Ignes-Mullol20}
J.~Ignés-Mullol, M.~Mora, B.~Martínez-Prat, I.~Vélez-Cerón, R.~S. Herrera,
  F.~Sagués, Stable and metastable patterns in chromonic nematic liquid
  crystal droplets forced with static and dynamic magnetic fields, Crystals
  10~(2) (2020) 138.

\bibitem{Blumler2023PracticalResonance}
P.~Bl{\"{u}}mler, H.~Soltner, {Practical Concepts for Design, Construction and
  Application of Halbach Magnets in Magnetic Resonance}, Applied Magnetic
  Resonance 54~(11-12) (2023) 1701--1739.
\newblock \href {https://doi.org/10.1007/s00723-023-01602-2}
  {\path{doi:10.1007/s00723-023-01602-2}}.

\bibitem{PhysRevResearch.5.033210}
R.~C.~V. Coelho, H.~Zhao, M.~Tasinkevych, I.~I. Smalyukh, M.~M. Telo~da Gama,
  Sculpting liquid crystal skyrmions with external flows, Phys. Rev. Res. 5
  (2023) 033210.
\newblock \href {https://doi.org/10.1103/PhysRevResearch.5.033210}
  {\path{doi:10.1103/PhysRevResearch.5.033210}}.

\bibitem{Coelho_2021}
R.~C.~V. Coelho, M.~Tasinkevych, M.~M.~T. da~Gama, Dynamics of flowing 2d
  skyrmions, J. Phys. Cond. Mat. 34~(3) (2021) 034001.
\newblock \href {https://doi.org/10.1088/1361-648x/ac2ca9}
  {\path{doi:10.1088/1361-648x/ac2ca9}}.

\bibitem{NAwa1978}
N.~Nawa, K.~Nakamura, Observation of forming process of bubble domain texture
  in liquid crystals, Japanese Journal of Applied Physics 17~(1) (1978)
  219--225.

\bibitem{Foster2019Two-dimensionalFerromagnets}
D.~Foster, C.~Kind, P.~J. Ackerman, J.~S.~B. Tai, M.~R. Dennis, I.~I. Smalyukh,
  {Two-dimensional skyrmion bags in liquid crystals and ferromagnets}, Nature
  Physics 15~(7) (2019) 655--659.
\newblock \href {https://doi.org/10.1038/s41567-019-0476-x}
  {\path{doi:10.1038/s41567-019-0476-x}}.

\bibitem{deSouza2025}
C.~C. de~Souza~Silva, M.~V. Correia, J.~.~P. Velasquez, Emergent
  self-propulsion of skyrmionic matter in synthetic antiferromagnets, Physical
  Review Letters 135~(8) (2025).

\bibitem{Leonov2014}
A.~O. Leonov, I.~E. Dragunov, U.~K. Rößler, A.~N. Bogdanov, Theory of
  skyrmion states in liquid crystals, Physical Review E - Statistical,
  Nonlinear, and Soft Matter Physics 90 (10 2014).
\newblock \href {https://doi.org/10.1103/PhysRevE.90.042502}
  {\path{doi:10.1103/PhysRevE.90.042502}}.

\bibitem{Oswald2023}
P.~Oswald, G.~Poy, J.~Ignés-Mullol, Tic reorientation under electric and
  magnetic fields in homeotropic samples of cholesteric lc with negative
  dielectric anisotropy, Crystals 13~(6) (2023) 957.

\bibitem{Sohn2019b}
H.~R.~O. Sohn, C.~D. Liu, Y.~Wang, I.~I. Smalyukh, Light-controlled skyrmions
  and torons as reconfigurable particles, Optics Express 27~(20) (2019).

\bibitem{Sohn2020}
H.~R.~O. Sohn, C.~D. Liu, R.~Voinescu, Z.~Chen, I.~Smalyukh, Optically enriched
  and guided dynamics of active skyrmions, Opt Express 28~(5) (2020)
  6306--6319.

\bibitem{Alvim2024}
T.~Alvim, M.~M. da~Gama, M.~Tasinkevych, Collective variable model for the
  dynamics of liquid crystal skyrmions, Communications Physics 7 (12 2024).
\newblock \href {https://doi.org/10.1038/s42005-023-01486-5}
  {\path{doi:10.1038/s42005-023-01486-5}}.

\end{thebibliography}





\end{document}